# Measuring radiofrequency fields in NMR spectroscopy using offset-dependent nutation profiles

Ahallya Jaladeep[1,*], Claris Niya Varghese[1,*] and Ashok Sekhar[1]

[1]Molecular Biophysics Unit, Indian Institute of Science, Bangalore 560 012, India

*Both these authors contributed equally to this work

**Corresponding author**:

Ashok Sekhar

Molecular Biophysics Unit

Indian Institute of Science Bangalore

Bengaluru - 560 012

Karnataka

India

Email: ashoksekhar@iisc.ac.in

**Abstract**

The application of NMR spectroscopy for studying molecular and reaction dynamics relies crucially on the measurement of the magnitude of radiofrequency (RF) fields that are used to nutate or lock the nuclear magnetization. Here, we report a method for measuring RF field amplitudes that leverages the intrinsic modulations observed in offset-dependent NMR nutation profiles of small molecules. Such nutation profiles are exquisitely sensitive to the magnitude of the RF field, and $B_1$ values ranging from 1-2000 Hz, as well the inhomogeneity in $B_1$ distributions, can be determined with high accuracy and precision using this approach. In order to measure $B_1$ fields associated with NMR experiments carried out on protein or nucleic acids, where these modulations are obscured by the large transverse relaxation rate constants of the analyte, our approach can be used in conjunction with a suitable external small molecule standard, expanding the scope of the method for large biomolecules.

## 1. Introduction

The NMR toolbox for characterizing molecular dynamics has diversified considerably in the last two decades with the introduction of methods such as chemical exchange saturation transfer (CEST) and relaxation dispersion (RD) (1-12). This has led to the detection and structural characterization of a number of sparsely populated biomolecular conformations implicated in enzyme catalysis (13, 14), molecular recognition (15-18) and protein folding (19, 20), as well as in aggregation (21, 22) and disease (23-25). While $R_{1\rho}$ RD experiments have identified transient Hoogsteen base pairs in duplex DNA (26, 27), mechanisms underlying base misincorporation during DNA replication (14, 28) and invisible excited states of HIV-1 TAR RNA (29), CEST has been employed to visualize higher energy functional conformations of Abl kinase (25), superoxide dismutase (23, 24) and the fluoride riboswitch (17), as well as to define the conformational selection mechanism underlying Hsp70 chaperone-substrate interactions (30). Moreover, the populations and lifetimes of transiently populated reaction intermediates in organic and metallorganic chemistry have recently been estimated with the CEST approach (31-35).

In both CEST and $R_{1\rho}$ RD experiments, an accurate measurement of the radiofrequency (RF) field strength is essential for quantifying thermodynamic, kinetic and structural parameters of the conformational exchange event (11, 36, 37). Over the years, a number of methods have been developed for determining the amplitude of the $B_1$ field, beginning with the sideband strategy outlined by Bloch (38) and demonstrated by Anderson in 1956 (39). With the advent of pulsed NMR, nutation resulting from on-resonance irradiation was proposed as an efficient method for calibrating the applied RF field (40, 41). In heteronuclear NMR involving $^{15}N$ or $^{13}C$ isotope-labeled samples, the magnitude $B_1$ field can be measured from the residual splitting observed in a multiplet pattern when the decoupling of the scalar-coupled nucleus is applied off-resonance (36, 42-44). While this method is ideal for $B_1$ amplitudes comparable to or larger than the heteronuclear coupling constant (90 - 150 Hz), much smaller $B_1$ fields are used routinely used in CEST experiments.

The current method for measuring weak $B_1$ fields proposed by Guenneugues *et al*. (45) is a variation of the nutation experiment (41, 46) where the RF field of desired strength is applied on z-magnetization for a systematically incremented time duration. Transverse components of magnetization are subsequently dephased with a gradient and the z-component is quantified through a read-pulse. Fourier transformation of this time-dependent signal provides both the amplitude and the probability distribution of the $B_1$ field across the sample. While this approach has been successful over a broad range of $B_1$ field strengths, it is challenging to use when measuring small RF fields of the order of 1 - 10 Hz. This is because of the need to place the RF transmitter on-resonance to within a value much smaller than the magnitude of the $B_1$ field, or alternatively to treat chemical shift offset as a fitting parameter while modeling intensities in the nutation spectrum.

In this manuscript, we report a method for measuring radiofrequency (RF) fields that makes use of modulations observed in the constant-time offset-dependent nutation profile of z-magnetization (CONDENZ) under the influence of RF radiation. This method enables the precise and accurate calibration of RF fields and is particularly useful for weak RF fields of the order of 1-10 Hz employed in CEST experiments. In addition, the CONDENZ approach provides a robust estimate of the inhomogeneity in the $B_1$ field that is required for the quantitative analysis of CEST and $R_{1\rho}$ RD profiles.

## 2. Materials and Methods

*2.1. Sample preparation*

An NMR sample of 100 mM unlabeled sucrose was prepared in 25 mM tris buffer at pH 7 with 90% $D_2O$/10% $H_2O$, 0.03 % $NaN_3$ and 1 mM DSS. This sucrose sample was used for all data collection for the $^{13}C$ CONDENZ profiles shown in Figure 1 and Figure S1. $^{13}C$ RF field strengths were also measured on a sample containing a mixture of 100 mM unlabeled sucrose, 1 mM methyl-$^{13}C$ α-ketobutyric acid and 50 mM benzaldehyde prepared in 25 mM tris buffer at pH 7 with 90% $D_2O$/10% $H_2O$, 0.03 % $NaN_3$ and 1 mM DSS.

RF fields on the $^{15}N$ nucleus were measured on two different NMR samples. The first one contained 1 mM $^{15}N^{\epsilon}$-labeled tryptophan and 0.8 mM U-$^{15}N$ ubiquitin, prepared in 25 mM sodium phosphate buffer at pH 7.0 with 25 mM NaCl, 1 mM EDTA, 0.03% $NaN_3$ and 2.5 % $d_6$-DMSO. The second one was used as an external standard for $^{15}N$ $B_1$ calibration and contained 1 mM $^{15}$N$^\varepsilon$-labelled tryptophan prepared in the same buffer as the $^{15}$N$^\varepsilon$-labeled tryptophan/U-$^{15}$N ubiquitin sample mentioned above. The aprotic solvent d$_6$-DMSO served as the lock solvent in both samples in order to eliminate artifacts from H/D solvent exchange (47).

U-$^{15}$N ubiquitin was overexpressed as a construct without any purification tag in *Escherichia coli* (*E.coli*) BL21(DE3) cells grown in 2xM9 media (48) with 1 g/L of $^{15}$NH$_4$Cl as the sole nitrogen source. Cell pellets were lysed by sonication and purified as described earlier (49). Briefly, the pH of the clarified cell lysate was adjusted to 4.5 by drop-wise addition of acetic acid to precipitate many of the endogenous *E.coli* proteins. The mixture was centrifuged at 15000 rpm for 1 hour and the supernatant was dialysed against 50 mM acetic acid/sodium acetate buffer at pH 4.5. Ubiquitin was then loaded on an SP-sepharose cation exchange column and eluted with a 0 - 500 mM NaCl gradient in the same buffer. Ubiquitin eluted at an NaCl concentration of 165 mM. Fractions containing ubiquitin were pooled, concentrated and loaded on a 16 x 60 Superdex 75 size exclusion chromatographic column for subsequent purification. Pure fractions were pooled, concentrated, aliquoted, flash frozen and stored at -80 °C.

*2.2. NMR spectroscopy*

All NMR experiments were performed at 25 °C on a 700 MHz Bruker spectrometer equipped with a room temperature triple resonance single-axis gradient TXI probe.

*2.3. CONDENZ measurements*

CONDENZ data were acquired in pseudo-2D mode using the pulse sequence shown in Figure 2. In each dataset, the RF field is positioned at a specific offset value and a 1D spectrum is acquired that typically contains only one peak at the chemical shift of proton which is scalar coupled to the target $^{13}$C ($^{15}$N) nucleus. The offset position of the RF field is then swept through the $^{13}$C ($^{15}$N) chemical shift of the peak of interest, generating one 1D spectrum corresponding to each offset. Every pseudo-2D CONDENZ dataset also contains a reference 1D spectrum acquired with the same pulse sequence, but with $t_{nut}$ set to 0. The sweep range and the offset step-size depend on the magnitude of the applied RF field and are tabulated in Table S1. For example, for a 10 Hz B$_1$ field, a pseudo-2D dataset containing 92 1D spectra was generated by sweeping -90 and 90 Hz using a uniform spacing between successive offsets of 2 Hz. Signal averaging over 16 transients for each 1D spectrum, corresponding to an average acquisition time of ∼ 40 min per $B_1$ field, usually gave sufficient signal-to-noise (SNR) for quantitative analysis. The relaxation delay used in all CONDENZ measurements was 1.5 s.

*2.4. Analysis of CONDENZ datasets*

Pseudo 2D CONDENZ data were processed and analysed using the Bruker Topspin version 4.0.7 software package. The 2D datasest was first split into individual 1D spectra using the command 'splitser', following which the peaks in 1D spectra acquired at different offsets were picked using the automated Topspin routine 'pps'. The intensities in the offset-dependent 1D spectra (I) were plotted as a ratio against the corresponding intensity in the reference spectrum ($t_{nut} = 0$, $I_0$) as a function of the offset at which the RF field is applied to generate CONDENZ profiles. Offsets were measured as differences from the on-resonance chemical shift of the target $^{13}$C nucleus, which was set to 0 Hz.

The errors in the intensity values were obtained using the 'SINO' routine. First, the regions corresponding to the signal and to noise in each individual 1D spectrum are selected. SINO is then used to determine the SNR for each spectrum in the CONDENZ dataset. The noise, which is used as the error estimate in intensity, is then evaluated from the offset-specific intensity and SNR values. Typically, 0.05 ppm surrounding the peak of interest and 0.5 ppm in the region 0.5 - 1.0 ppm are chosen as the signal and noise regions in SINO. For offset values near resonance, the intensity is close to 0. The error value assigned in such cases is the average of the errors estimated from spectra belonging to the same pseudo-2D dataset where intensity is sufficiently large to quantify the error.

## 3. Results and Discussion

*3.1. Intensity modulations observed in offset-dependent nutation profiles*

Figure 1A and Figure S1 show CONDENZ nutation profiles of the anomeric $^{13}$C magnetization of the glucose ring in sucrose (referred to herewith as the sucrose anomeric carbon) for $B_1$ fields ranging from 1 - 2000 Hz. Nutation data were acquired using a selective 1D-based $^{13}$C-CEST pulse sequence shown in Figure 2. In this sequence, $^{13}$C z-magnetization of the anomeric sucrose carbon ($C_z$) is created from the single-bonded anomeric $^1$H magnetization by transfer via a selective J cross-polarization module (50, 51) (Fig. 2). $C_z$ is then subjected to a $B_1$ field applied at a specific $^{13}$C offset for a constant nutation time $t_{nut}$, during which scalar coupled protons are decoupled via a DIPSI-2 (52) composite pulse decoupling train. The $^{13}$C z-magnetization at the end of $t_{nut}$ is then transferred back to $^1$H using the same selective Hartmann-Hahn transfer for detection. A reference spectrum is acquired for each value of the $B_1$ field using the same pulse sequence but where the nutation period is absent. CONDENZ profiles in Figure 1A graph the intensity of the sucrose anomeric $^1$H peak (I) as a ratio against the intensity of the target peak in the reference spectrum ($I_0$) as a function of the $^{13}$C offset at which the $B_1$ field is applied.

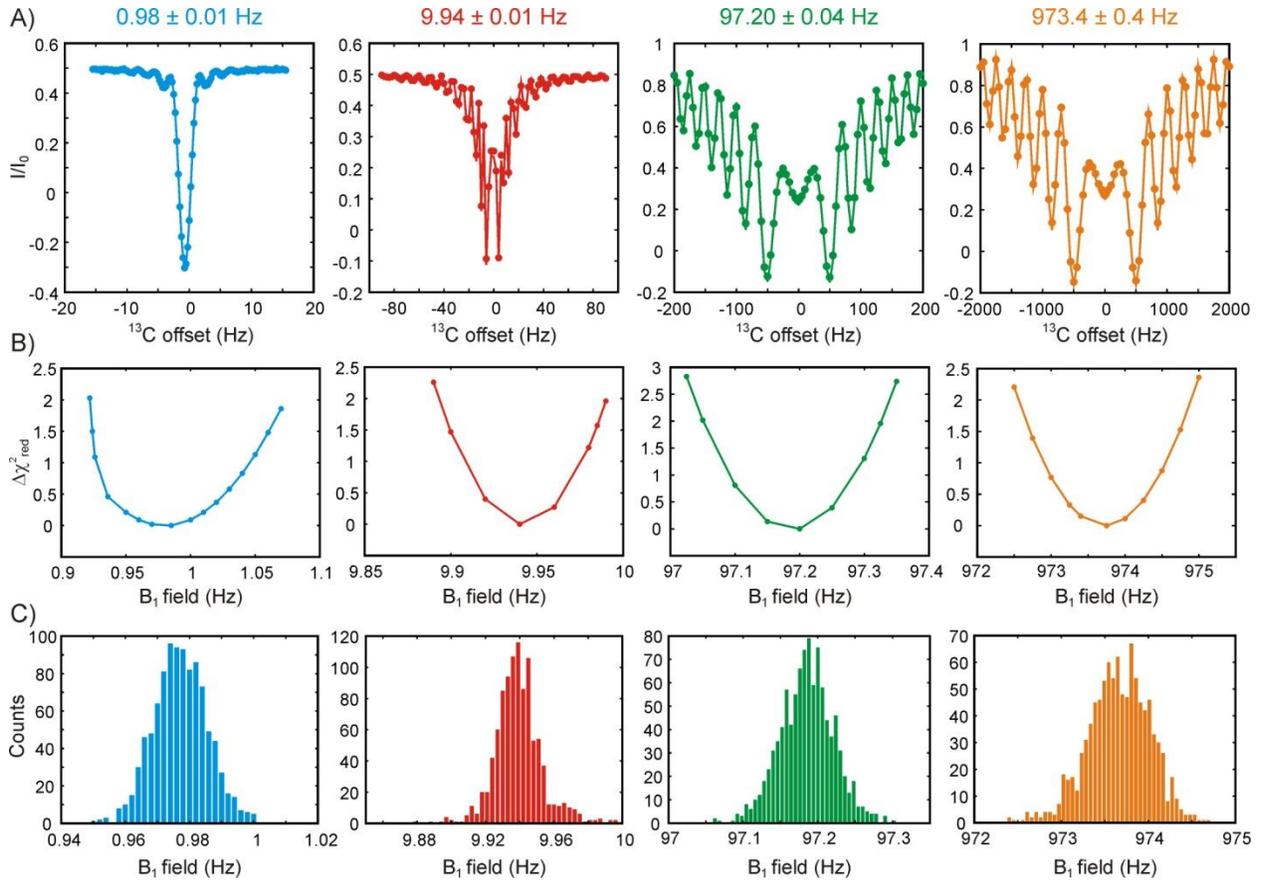

**Figure 1**. A) $^{13}$C CONDENZ profiles for RF amplitude settings of 1 Hz (cyan, $t_{nut}$ = 400 ms), 10 Hz (red, $t_{nut}$ = 400 ms), 100 Hz (green, $t_{nut}$ = 50 ms) and 1000 Hz (orange, $t_{nut}$ = 5 ms). Each profile is plotted as the intensity (I) of the sucrose anomeric $^1$H peak, normalized to the intensity in a reference spectrum acquired without $t_{nut}$ ($I_0$), as a function of the $^{13}$C offset at which the $B_1$ field is applied. Solid lines are fits of the data (circles) to the Bloch equations (Eq. 9). The $B_1$ estimate from the fit is indicated at the top of the plot along with the error obtained through a bootstrap routine. The $\chi^2_{red}$ surface for the fit, plotted as the increase in $\chi^2_{red}$ from the best fit value, (B) evaluated by keeping the $B_1$ field fixed at various values during the fitting routine, as well as the bootstrap distribution for each fit (C) are shown below each CONDENZ profile.

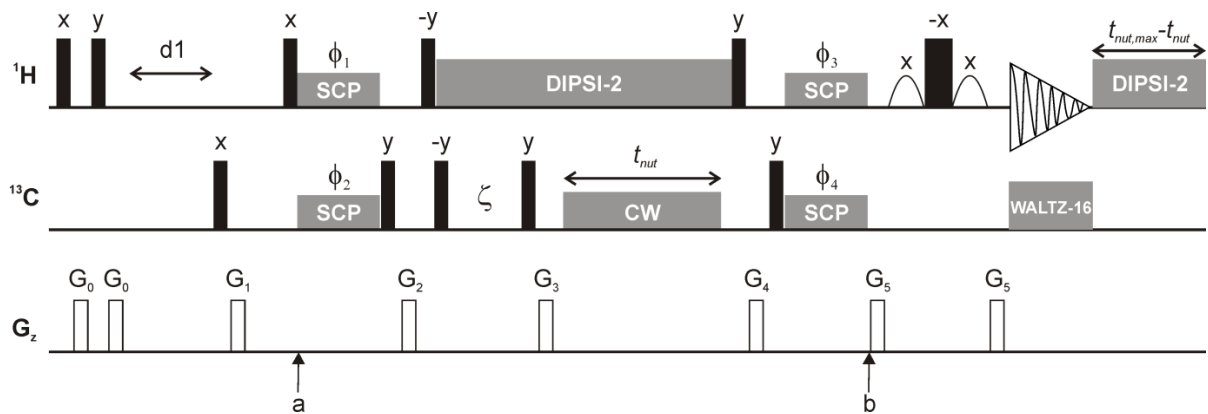

**Figure 2.** The selective $^{13}$C 1D-based pulse sequence used for acquiring $^{13}$C CONDENZ profiles. Hard 90° and 180° pulses on both $^1$H and $^{13}$C channels are depicted as black narrow and wide rectangles respectively and applied at the maximum available power. The $^1$H transmitter is positioned on-resonance to the target proton signal in-between points a and b, and on-resonance to water through the rest of the pulse sequence. The $^{13}$C transmitter is placed on-resonance to the target carbon throughout the pulse sequence except during the nutation period, when it is varied systematically to generate $^{13}$C offset-dependent intensities. In this pulse sequence, $^1$H magnetization is first destroyed by a $90°_x - G_z - 90°_y - G_z$ module and $^1$H z-magnetization is allowed to recover during the subsequent 1.5 s relaxation delay (d1). This module ensures that the same magnitude of $^1$H polarization is available at the beginning of each slice of a CONDENZ dataset. Following the d1 period, equilibrium $^{13}$C polarization is dephased with a gradient and $^1$H magnetization is transferred to $^{13}$C through a selective cross-polarization (SCP) element (51). SCP is achieved by application of matched RF fields with an amplitude of 130 Hz on the $^1$H and $^{13}$C channels for a period of 6.3 ms for AX and 4.6 ms for AX$_3$ spin systems. After magnetization transfer, $^{13}$C coherence is flipped onto the z-axis and residual transverse $^1$H and $^{13}$C components are destroyed with a gradient pulse. Subsequently, $^{13}$C transverse magnetization is created with a $90°_{-y}$ pulse and passed through a filter for eliminating residual magnetization

originating from resonances having similar $^{13}$C chemical shifts (within ~ 1 ppm). In this filter, transverse magnetization is maintained for a period $\zeta = \frac{1}{4\delta}$, where δ is the difference in chemical shifts (Hz) between the target $^{13}$C chemical shift and the second nucleus in its vicinity (53). In the ζ period, coherence from the on-resonance target $^{13}$C does not precess in the rotating frame and remains oriented along the -x axis, while the coherence from the second nucleus precesses by 90° to align along the y or the -y axis (depending on the sign of its $^{13}$C offset in the rotating frame). The application of a second $90°_y$ pulse places the target polarization along the z-axis but does not affect the coherence of the second nucleus, which is then dephased by a gradient before the $t_{nut}$ delay. $^1$H decoupling during both the ζ and $t_{nut}$ periods is implemented using a 4 kHz DIPSI-2 composite decoupling scheme (52). Following $B_1$ irradiation along the x-axis during $t_{nut}$, residual transverse $^{13}$C magnetization is again eliminated with a gradient pulse and the z-component is transferred to $^1$H through an SCP element. Water suppression is implemented using 1.5 ms rectangular water-selective pulses which are shown as open curves. $^{13}$C decoupling during acquisition is carried out using a WALTZ-16 train (54). A reference spectrum is acquired using the same sequence but lacking the $t_{nut}$ nutation period. In order to ensure that heating from the DIPSI-2 $^1$H decoupling is the same in the reference spectrum as well as in spectra acquired with different values of $t_{nut}$, a heat compensation element is inserted in the pulse sequence immediately after completion of data acquisition. During this heat compensation element, the 4 kHz DIPSI-2 $^1$H decoupling field is turned on for a period of $t_{nut,max}$ - $t_{nut}$, where $t_{nut,max}$ is the maximum nutation time. The complete phase cycling for this sequence is: $\phi_1$ = {y,y,y,y,y,y,y,y,-y,-y,-y,-y,-y,-y,-y,-y}, $\phi_2$ = {-x,x}, $\phi_3$ = {x,x,x,x,-x,-x,-x,-x}, $\phi_4$ = {x,x,-x,-x} and $\phi_R$ = {x,-x,-x,x,-x,x,x,-x,-x,x,x,-x,x,-x,-x,x}, but an 8-step phase cycle is sufficient. Gradient strengths are applied in the smooth square shape with the following strengths (% of maximum of ~ 50 G/cm) and durations: $G_0$ (11 %, 500 μs), $G_1$ (17 %, 500 μs), $G_2$ (83 %, 500 μs), $G_3$ (93 %, 500 μs), $G_4$ (71 %, 500 μs), $G_5$ (66 %, 300 μs). $^{15}$N CONDENZ data are acquired with the same pulse sequence but with $^{13}$C pulses replaced by $^{15}$N ones. The $B_1$ field strength and duration for SCP in this case are 90 Hz and 11 ms respectively.

## 3.2. Theoretical basis for modulations seen in CONDENZ profiles

The presence of offset-dependent modulations observed in CONDENZ profiles can be explained through an analysis of the Bloch equations. For a single-spin-1/2 system, the Bloch equations in the rotating frame for the three components of magnetization ($M_x$, $M_y$ and $M_z$) in the presence of a RF field of amplitude $\omega_1$ (rad/s) applied along the x-axis are given by (55):

$$\frac{d\vec{M}}{dt} = R\vec{M} \qquad (1)$$

where $\vec{M} = \begin{bmatrix} E & M_x & M_y & M_z \end{bmatrix}^T$ ($T$ being the transpose operation), $\Delta$ is the chemical shift offset (rad/s) and

$$R = \begin{bmatrix} 0 & 0 & 0 & 0 \\ 0 & -R_2 & -\Delta & 0 \\ 0 & \Delta & -R_2 & -\omega_1 \\ R_1 M_0 & 0 & \omega_1 & -R_1 \end{bmatrix}. \qquad (2)$$

The formal solution to this set of coupled differential equations is:

$$\vec{M}(t) = e^{Rt}\vec{M}(0). \qquad (3)$$

For small molecules where relaxation rates are slow compared to $\Delta$ or $\omega_1$, we can neglect relaxation and the simplified equations of motion in the rotating frame become:

$$\begin{aligned} \frac{dM_x(t)}{dt} &= -\Delta M_y(t) \\ \frac{dM_y(t)}{dt} &= \Delta M_x(t) - \omega_1 M_z(t) \\ \frac{dM_z(t)}{dt} &= \omega_1 M_y(t) \end{aligned} \qquad (4)$$

Assuming initial conditions of:

$$\begin{aligned} M_x(0) &= M_y(0), \\ M_z(0) &= M_0 \end{aligned}, \qquad (5)$$

the analytical solution for $M_z(t_{nut})$ is given by:

$$\frac{M_z(t_{nut})}{M_0} = 1 - \left[\frac{\omega_1^2 t_{nut}^2}{2}\left(\frac{\sin\phi}{\phi}\right)^2\right], \qquad (6)$$

where

$$\phi = \frac{\sqrt{\Delta^2 + \omega_1^2}\, t_{nut}}{2} \qquad (7)$$

The trajectory of z-magnetization as a function of the offset $\Delta$ is thus a squared sinc function, matching experimental observations shown in Figure 1A. The argument of the sinc function depends on the variable chemical shift offset $\Delta$, as well as the values of the nutation time and the amplitude of the applied radiofrequency field, which are held constant while acquiring nutation profiles.

*3.3. CONDENZ profiles are exceptionally sensitive to the $B_1$ field strength*

Interestingly, simulations of the Bloch equations (Fig. 3) reveal that these nutation profiles are exquisitely sensitive to the magnitude of the RF field; for example, profiles simulated with $B_1$ values of 3 and 3.2 Hz are visibly different. Therefore, we reasoned that experimentally acquired nutation profiles should enable us to measure the corresponding $B_1$ field with high precision.

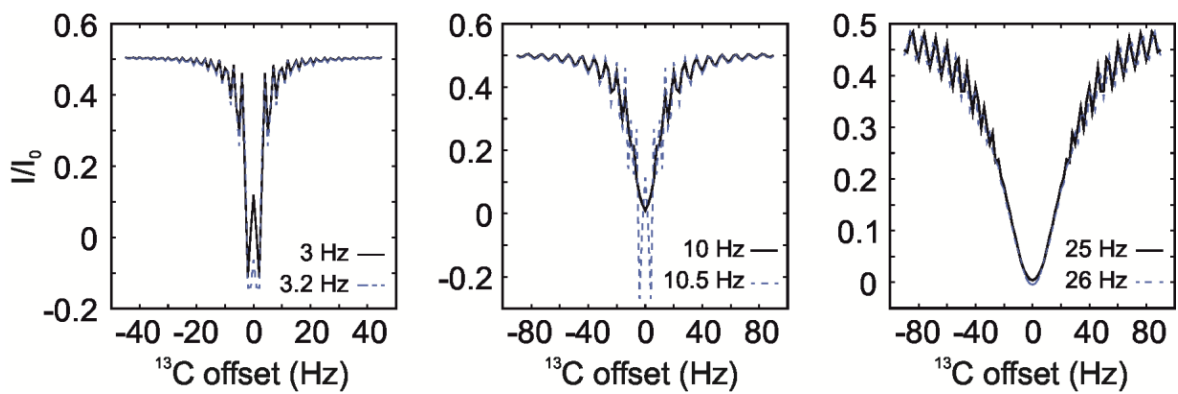

**Figure 3.** CONDENZ profiles simulated for a single $^{13}$C spin for pairs of $B_1$ fields (3 Hz, 3.2 Hz, left), (10 Hz, 10.5 Hz, middle) and (25 Hz, 26 Hz, right) demonstrating the high sensitivity of these profiles to small changes in $B_1$ field amplitude. Datasets were generated using the following parameters: $t_{nut} = 0.4$ s, $R_1 = 1.7$ s$^{-1}$, $R_2 = 2.0$ s$^{-1}$ and $I_{B1} = 5$ %.

Accordingly, we modeled nutation profiles shown in Figure 1A using the Bloch equations, assuming the anomeric $^{13}$C to be a single-spin-1/2 system. Longitudinal ($R_1$) and transverse ($R_2$) relaxation rates of the anomeric carbon, as well as its resonance offset and the $B_1$ field were treated as fitting parameters. The RF amplitude was assumed to follow a Gaussian distribution with its standard deviation defined as the $B_1$ inhomogeneity ($I_{B1}$) (56). N (typically, N = 11) samples of the $B_1$ field ($B_{1s}$) were drawn from this distribution within the range $\left[-\left(B_1+2I_{B1}\right),\left(B_1+2I_{B1}\right)\right]$ and the probabilities of their occurrence were determined from the equation for the Gaussian probability distribution function as:

$$P(B_{1s}) = \frac{1}{\sqrt{2\pi I_{B1}^2}} \exp\left(-\frac{(B_{1s}-B_1)^2}{2I_{B1}^2}\right) \qquad (8)$$

The formal solution of the Bloch equations depends on the particular value of $B_{1s}$, so that the overall magnetization at the end of the time-evolution is a weighted sum of the individual evolutions, with the weights given by $P(B_{1s})$ as:

$$\vec{M}(t) = \sum_N P(B_{1S}) M_S(t) = \sum_N P(B_{1S}) e^{R(B_{1S})t} \vec{M}(0) \qquad (9)$$

where the evolution matrix R now is a function of the particular sampled value of $B_{1S}$.

Nutation profiles can be fit very well using the single-spin-1/2 Bloch equations described by Eq. 9 to recover the magnitude of the RF field (Fig. 1A, Fig. S1). $\chi_{red}^2$ surfaces as a function of $B_1$ are very steep (Fig. 1B, Fig. S1), demonstrating the robustness of the $B_1$ values obtained from data fitting. In order to determine the precision of the measured $B_1$ values, we estimated errors using a bootstrap algorithm (57), where 1000 bootstrapped datasets for each $B_1$ field were constructed from the nutation profile by random sampling with replacement and fit to the Bloch equations. The resulting bootstrapped $B_1$ distributions (Fig. 1C, Fig. S1, Table 1) are narrow with standard deviations ranging from 0.02 - 1% of the measured $B_1$ value, clearly showing that highly precise $B_1$ measurements can be made from the modeling of offset-dependent nutation curves.

| $B_{1\ setting}$ (Hz) | $B_{1\ CONDENZ}$ (Hz) | $I_{B1}$ (%) | $B_{1\ nutation}$ (Hz) |
|---|---|---|---|
| 1 | 0.98 ± 0.01 | ND | 0.91 |
| 2 | 2.00 ± 0.01 | ND | 2.04 |
| 3 | 2.95 ± 0.01 | ND | 3.03 |
| 5 | 4.95 ± 0.04 | ND | 5.03 |
| 7.5 | 7.47 ± 0.02 | 5.3 ± 0.5 | 7.50 |
| 10 | 9.94 ± 0.01 | 5.0 ± 0.2 | 9.98 |
| 15 | 14.98 ± 0.02 | 4.6 ± 0.3 | 14.99 |
| 20 | 19.79 ± 0.03 | 4.2 ± 0.2 | 20.07 |
| 25 | 24.99 ± 0.03 | 4.1 ± 0.2 | 24.98 |
| 30 | 29.67 ± 0.04 | 4.6 ± 0.3 | 30.04 |
| 100 | 97.20 ± 0.04 | 4.65 ± 0.04 | 96.30 |
| 500 | 486.3 ± 0.1 | 5.42 ± 0.05 | 485.9 |
| 750 | 731.6 ± 0.2 | 5.21 ± 0.04 | 731.1 |
| 1000 | 973.4 ± 0.4 | 4.72 ± 0.03 | 975.6 |
| 1500 | 1456.9 ± 0.3 | 5.16 ± 0.04 | 1465.0 |
| 2000 | 1943 ± 3 | 6.0 ± 0.1 | 1965 |

**Table 1**. A comparison of $^{13}$C RF field amplitudes measured on a sucrose sample using the on-resonance nutation experiment ($B_{1\ nutation}$) and CONDENZ ($B_{1\ CONDENZ}$). $B_{1\ nutation}$ values are reported to the same number of decimal places as the $B_{1\ CONDENZ}$ values to facilitate comparison. $B_1$ inhomogeneities extracted from the CONDENZ profiles at each $B_1$ field are listed in column 4. (ND: not determined)

*3.4. Evaluating the accuracy of $B_1$ amplitudes extracted from CONDENZ profiles*

As the next step, we evaluated the data for the presence of systematic errors that could bias the $B_1$ estimates. First, we experimentally measured the magnitude of $B_1$ using a second method proposed by Guenneugues *et al* (45, 58), in which magnetization is nutated by an external $B_1$ field applied on-resonance to the peak of interest for a variable nutation time. The time-dependent signal intensity is then Fourier transformed to determine the $B_1$ field. $B_1$ values obtained from CONDENZ profiles and the on-resonance nutation method agree very well with an $R^2$ value of 0.99 (Fig. 4, Table 1), demonstrating the reliability of the RF field measurements that can be made with the CONDENZ method.

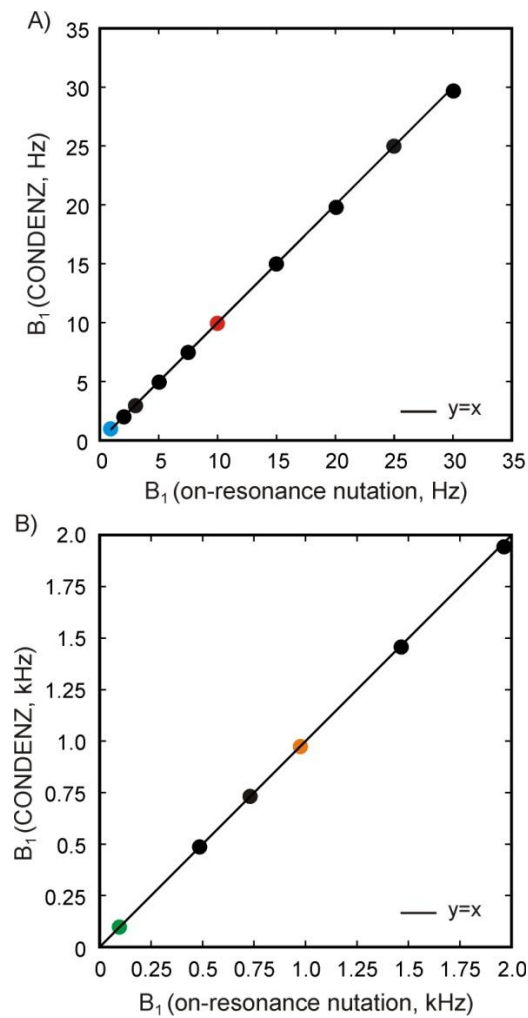

**Figure 4**. A comparison of the $B_1$ field amplitude obtained by modeling CONDENZ profiles (y-axis) with the values determined from an on-resonance nutation experiment (x-axis) for weak (1-35 Hz, panel A) and strong (100 - 2000 Hz, panel B) $B_1$ fields. The solid line in both panels is a y=x function. $B_1$ data points obtained by fitting CONDENZ profiles shown in Figure 1 are indicated with the same colour scheme as in Figure 1. Error bars are generally smaller than the size of the data points and not visible in the plot.

Second, we addressed the possibility of systematic deviations originating from the use of single-spin-1/2 Bloch equations for modeling nutation profiles. While the anomeric proton in sucrose is decoupled from the covalently bonded carbon ($^1J_{CH}$ = 169 Hz) during $t_{nut}$ through the use of a 4 kHz DIPSI-2 composite pulse decoupling scheme, the anomeric carbon is also scalar coupled to the proton on the neighbouring carbon that is off-resonance to the DIPSI-2 field by 1.86 ppm, through a two-bond $^2J_{CH}$ of 7 Hz (59). These two-bond and three-bond couplings could impact the fit $B_1$ values especially at small amplitudes of the RF field. In order to probe this possibility, we first simulated nutation profiles at various $B_1$ fields using product operators for an AMX spin system, where A and M are the anomeric carbon and proton ($^1J_{CH}$ = 169 Hz) that are on-resonance to the $B_1$ and decoupling fields respectively, and X is the neighbouring proton that is off-resonance to the decoupling field and scalar-coupled to the anomeric carbon with a $^2J_{CH}$ of 7 Hz and to the anomeric proton with a $^3J_{HH}$ of 4 Hz. The simulated profiles were then fit with the single-spin-1/2 equations above to extract the amplitude of the $B_1$ field (Fig. S2). The input and fit $B_1$ values correlate excellently ($R^2$ = 0.99) and agree to within 0.5 % over a range of $B_1$ values from 0.5 - 10 Hz (Fig. S2, Table S2), confirming that the use of the single-spin-1/2 equations for modeling the nutation profiles does not systematically distort the measured $B_1$ values.

*3.5. CONDENZ profiles provide robust estimates of the RF inhomogeneity*

Having established the utility of our method for accurately and precisely measuring RF fields, we next asked whether the offset-dependent modulations observed in our experiments are sensitive to $B_1$ inhomogeneity. In order to address this question, we constructed $\chi^2_{red}$ curves that reflect the robustness of the parameter estimates of $I_{B1}$. These curves show pronounced minima for $B_1$ > 5 Hz (Fig. 5A, Fig. S1), confirming that $I_{B1}$ can also be measured reliably with our method. The errors in $I_{B1}$ range from ~ 9 % at 7.5 Hz to < 1% at 1 kHz (Table 1), showing that the estimates are more precise at higher RF field strengths. For $B_1 \leq 5$ Hz, $\chi^2_{red}$ curves are often flat on the lower limit of $I_{B1}$, suggesting that only an upper estimate can be extracted reliably (Fig. S1). We verified these conclusions by fitting data simulated using the procedure detailed above to Eq. 9. $I_{B1}$ values recovered from the fit match very well with the input values to within 0.5 % for $B_1$ > 5 Hz (Table S2), while there is a systematic difference of the order of 5 % between input and fit values for $B_1 \leq 5$ Hz.

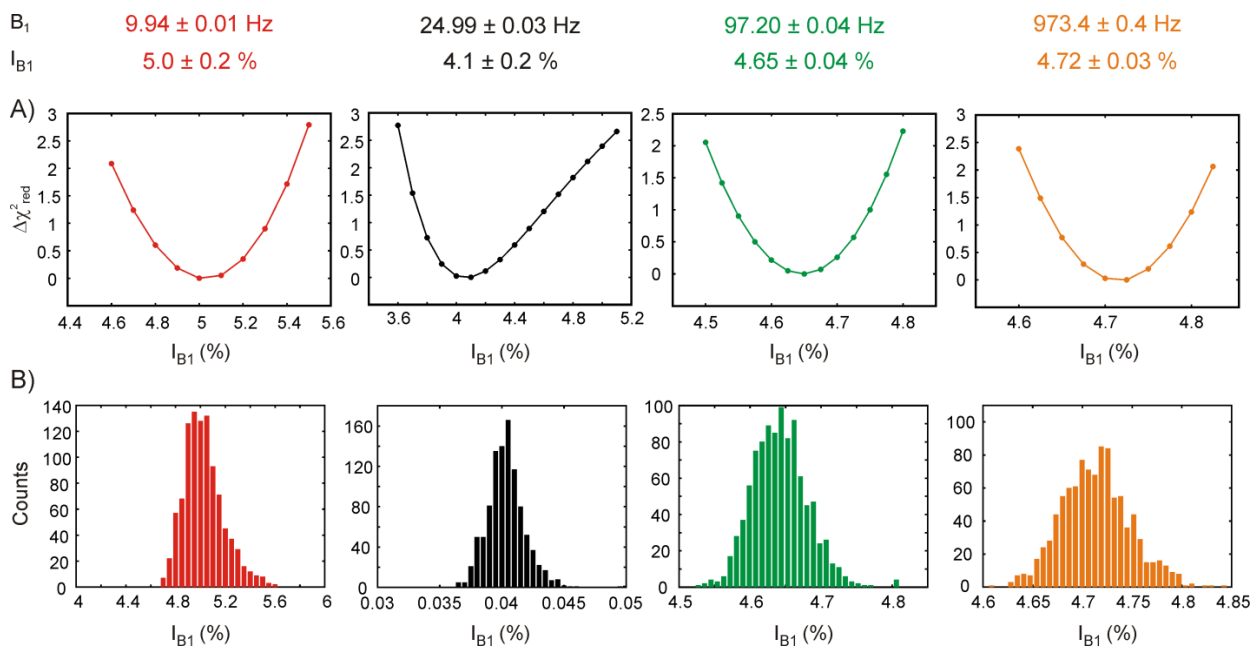

**Figure 5**. $\chi^2_{red}$ surfaces (A) and bootstrap (B) distributions for $B_1$ inhomogeneity ($I_{B1}$) evaluated by modeling CONDENZ profiles. The inhomogeneity is depicted as a percentage of the $B_1$ field. The values of $B_1$ and $I_{B1}$ obtained from the CONDENZ profiles are indicated at the top of the figure. $B_1$ measurements made from profiles shown in Figure 1 are coloured with the same scheme.

*3.6. Estimating $^{15}$N RF amplitudes using CONDENZ data*

Since $R_{1\rho}$ and CEST data are acquired primarily on $^{13}$C and $^{15}$N nuclei (36), we next explored the possibility of measuring $^{15}$N RF fields using the CONDENZ approach. We chose $^{15}$N$^\varepsilon$-labeled Trp as a suitable small molecule because of a number of favorable properties such as easy availability, the slow solvent exchange rate of the indole $^1$H$^\varepsilon$ (60), a sharp indole $^1$H$^\varepsilon$-$^{15}$N$^\varepsilon$ correlation, and a $^{15}$N chemical shift that falls within the resonance frequency range of typical protein and nucleic acid molecules. In order to eliminate potential interference from H/D exchange in the nutation profiles, we used 2.5 % DMSO-d$_6$ as the lock solvent (47). CONDENZ profiles of the Trp indole $^{15}$N$^\varepsilon$ nucleus also show offset-dependent modulations in the presence of a $B_1$ field similar to those seen for the sucrose anomeric $^{13}$C (Fig. 6A, Fig. S3), confirming that these modulations are independent of the identity of the nucleus. Nutation profiles were modeled using the Bloch equations (Eq. 9) to extract values of $B_1$ and $I_{B1}$. Similar to $^{13}$C, the amplitude of $B_1$ fields applied on the $^{15}$N channel also can be measured accurately and precisely with steep $\chi^2_{red}$ and small errors (0.05 - 0.5 %) (Fig. S3, Table S3) for $B_1$ fields larger than 2 Hz, demonstrating the generality of the methodology. In contrast, there is a 6 % difference between the $B_1$ fields measured using the on-resonance and CONDENZ approaches for $B1 \leq 2$ Hz, highlighting the difficulty of locating the exact $^{15}$N chemical shift of the Trp indole $^{15}$N$^\varepsilon$ while acquiring on-resonance nutation data.

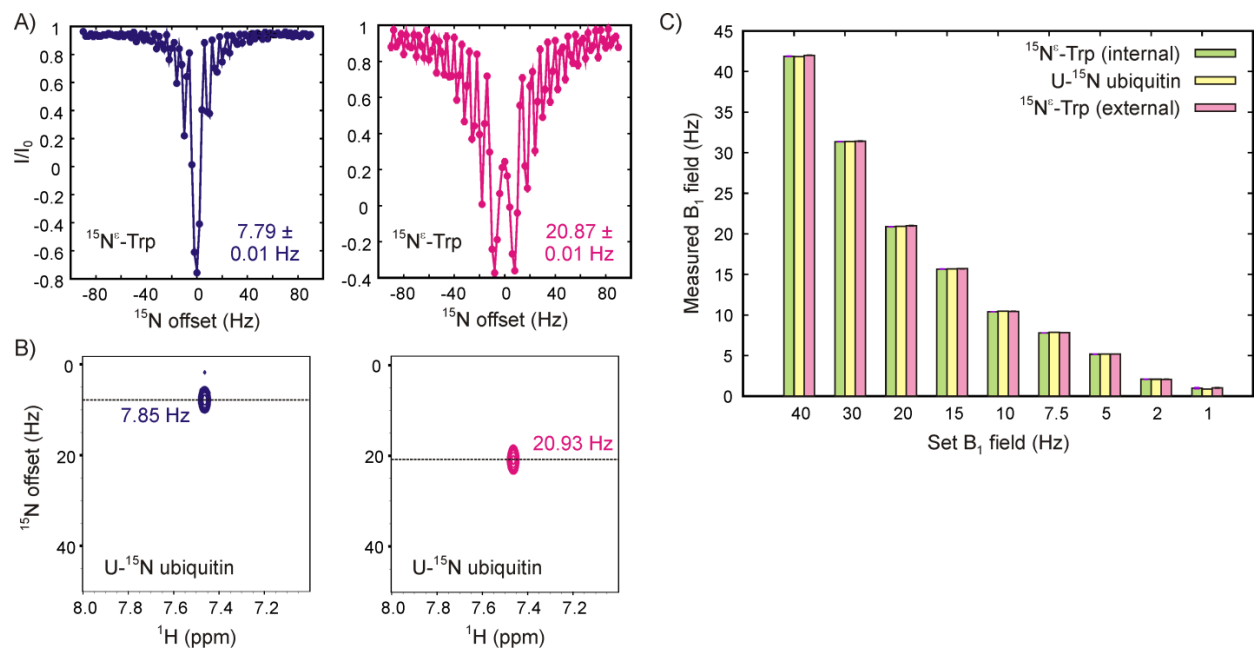

**Figure 6**. A) $^{15}$N CONDENZ profiles for $B_1$ settings of 7.5 Hz (blue, left, $t_{nut}$ = 200 ms) and 20 Hz (magenta, right, $t_{nut}$ = 200 ms), plotted as the intensity ratio of the indole H$^\varepsilon$ resonance in spectra acquired with (I) and without $t_{nut}$ (I$_0$) against the $^{15}$N offset at which the $B_1$ field is applied. Solid lines are fits of the data to the Bloch equations (Eq. 9). The $B_1$ values obtained by modeling the CONDENZ profiles and the corresponding standard deviations recovered from a bootstrapping procedure are indicated at the top of the plot. Data were acquired on a sample containing both $^{15}$N$^\varepsilon$-Trp (internal standard) and U-$^{15}$N ubiquitin. B) On-resonance nutation spectra acquired using the pulse sequence of Guenneugues *et al.* (45, 58) on a peak belonging to $^{15}$N-labeled ubiquitin for a $B_1$ setting of 7.5 Hz (blue, left) and 20 Hz (magenta, right). The same sample as for panel A was used in these experiments. C) Histogram depicting the comparison between $B_1$ field strengths measured on U-$^{15}$N ubiquitin (yellow), an internal $^{15}$N$^\varepsilon$-Trp standard (green) and an external $^{15}$N$^\varepsilon$-Trp standard (pink) for $B_1$ amplitude settings ranging from 1 - 40

Hz, showing excellent agreement between the three values for all $B_1$ fields. Error bars are of the order of the line thickness and not readily visible.

*3.7. Practical aspects of using CONDENZ profiles to determine $B_1$ field strength*

There are a few practical considerations that govern the utilization of this methodology. First and foremost, the nutation parameters such as $t_{nut}$ and the offset spacing ($\delta\Delta$) must be adjusted so that the squared-sinc modulation is clearly observed in the CONDENZ profile. If $t_{nut}$ or $\delta\Delta$ are too large, or if the signal-to-noise ratio (SNR) is too low so that clear modulations are not observed, the $B_1$ field cannot be extracted reliably. Parameters that provide tractable nutation profiles and accurate $B_1$ fields in the range from 1 - 2000 Hz are listed in Table S1 and can be used as initial estimates, though we have observed that $t_{nut}$ values may have to be slightly modified based on the $R_2$ of the observed nucleus; if the $R_2$ is higher than for the anomeric carbon of sucrose, smaller $t_{nut}$ values can be employed to visualize the modulations clearly. Second, a vast number of CEST and $R_{1\rho}$ experiments are carried out on large biomolecules such as proteins and nucleic acids, where $R_2$ is too large to observe squared-sinc modulations. In such cases, we asked whether small molecules can be used as internal or external standards for measuring the $B_1$ field. First, we doped a $^{15}$N-labeled ubiquitin sample with an internal $^{15}N^{\varepsilon}$-Trp standard and measured the RF field amplitude for $B_1$ values ranging from 1 - 50 Hz using the on-resonance nutation experiment for ubiquitin and the CONDENZ approach for Trp (Fig. 6B). The values agree to within 1 % and an $R^2$ of 0.99 (Table S4), demonstrating that both molecules experience the same $B_1$ field. Next, we made an external standard of $^{15}N^{\varepsilon}$-Trp in the same buffer and measured the $B_1$ field using CONDENZ. The magnitude of the RF field matches excellently with those estimated from the internal Trp standard as well as with ubiquitin (to within 1 %), provided the tuning, matching, pulse widths and power levels are left unchanged between the analyte and the external standard (Table S4). These results unequivocally show that the CONDENZ approach for determining the RF amplitude can be used in conjunction with an external small molecule standard and extends its utility for sensitive biomolecular samples that may be affected by the addition of small molecule standards. Third, $R_{1\rho}$ and CEST experiments on biomolecules span a wide range of resonance offsets (28, 29, 56, 61-68). Specifically, experiments on $^{13}$C nuclei in proteins are acquired on moieties with chemical shifts from 5 ppm ($^{13}C^{\delta}$ of Ile) to 175 ppm

(carbonyl carbons) that corresponds to 34 kHz on an 800 MHz spectrometer. In order to determine whether the same external standard (eg. the anomeric $^{13}$C of sucrose resonating at 92 ppm) can be used to estimate the $B_1$ amplitude for such a wide range of $^{13}$C transmitter frequencies, we prepared a sample containing a mixture of benzaldehyde, sucrose and α-ketobutyric acid, and acquired CONDENZ profiles on the methyl-$^{13}$C of $^{13}$CH$_3$ α-ketobutyric acid ($\varpi$ = 6.5 ppm), the anomeric $^{13}$C of sucrose ($\varpi$ = 92 ppm), an aromatic $^{13}$C of benzaldehyde ($\varpi$ = 135.5 ppm) and the aldehydic $^{13}$C of benzaldehyde ($\varpi$ = 196.5 ppm). The magnitude of the $B_1$ field determined with these four $^{13}$C nuclei spanning a range of ~ 200 ppm in chemical shift agree to within a maximum deviation of 1.2 % for $B_1$ fields of 5 Hz or larger, while a deviation of 12 % is observed for the 2 Hz case (Table S6). Though the deviations observed in the mixture sample are small, these experiments suggest that it is preferable to use a calibration standard whose resonance frequency matches well with the nucleus on which $R_{1\rho}$ or CEST experiments are carried out, especially where small (1-3 Hz) $B_1$ fields are involved. Finally, a SNR of the reference standard of 200 is sufficient to guarantee good quality nutation profiles which can be analyzed to determine the amplitude of the RF field. For a 100 mM unlabeled sucrose sample (natural abundance $^{13}$C) or 1 mM $^{15}$N$^\varepsilon$-Trp, nutation data with such SNR can be obtained in 45 min on a 700 MHz spectrometer equipped with a room-temperature probe, underscoring the accessibility and ease of application of our method.

4. Conclusions

The CONDENZ approach is particularly useful for measuring weak $B_1$ fields of the order of 1-10 Hz that are routinely employed in CEST experiments. In this $B_1$ regime, the on-resonance nutation method is susceptible to interference from off-resonance effects, because locating the resonance frequency to within a few Hz is difficult for biomolecules with 10 Hz or larger linewidths. The CONDENZ method, however, is immune to off-resonance effects as the approach relies on measurements made by varying the chemical shift offset, the chemical shift offset is a fitting parameter and therefore does not have to be exactly identified. In addition, our experiments demonstrate that $B_1$ calibration can be carried out with a small molecule external standard, eliminating the necessity for finding an isolated protein or nucleic acid resonance that does not undergo conformational exchange for this purpose. We anticipate such flexibility to be particularly useful for large biomolecules and intrinsically disordered proteins, whose 2D correlation spectra are characterized by severe peak overlap.

**Acknowledgments**

We thank Dr. Tairan Yuwen for help with the simulations for the AMX spin system as well as Dr. Pramodh Vallurupalli and Dr. Siddhartha Sarma for a critical reading of the manuscript. This work was supported by the DBT/Wellcome Trust India Alliance Fellowship (grant no.: IA/I/18/1/503614) and a DST/SERB Core Research grant (no. CRG/2019/003457), as well as a start-up grant from IISc awarded to A.S. We also acknowledge funding for infrastructural support from the following programs of the Government of India: DST-FIST, UGC-CAS, and the DBT-IISc partnership program. C.N.V. thanks the Ministry of Human Resource Development, Government of India for fellowship support through the Prime Minister's Research Fellows scheme. A.J. thanks IISc Bangalore for fellowship support.

**Supporting Information**

Figures showing CONDENZ profiles for sucrose and Trp for all RF field strengths, simulations showing CONDENZ profiles for an AMX spin system, table detailing the fitted $B_1$ and $I_{B_1}$ values for Trp, a comparison between input $B_1$ field strength in the AMX simulation and the fit value assuming a single-spin approximation, tabulated values of RF amplitudes obtained with ubiquitin as well as internal and external small molecule standards, and a comparison of RF amplitudes obtained with different small molecule probes.

# Supporting Information

# Measuring radiofrequency fields in NMR spectroscopy using offset-dependent nutation profiles

Ahallya Jaladeep[1,*], Claris Niya Varghese[1,*] and Ashok Sekhar[1]

[1]Molecular Biophysics Unit, Indian Institute of Science, Bangalore 560 012, India
*Both these authors contributed equally to this work

# Supporting Figures

**Figure S1.** CONDENZ profiles (row 1) acquired on the anomeric carbon of sucrose at various $B_1$ fields ranging from 1 - 2000 Hz. Solid lines are fits of the data to the single-spin-1/2 Bloch equations (main manuscript, Eq. 9). The $\chi^2_{red}$ surface for each fit, evaluated by keeping the $B_1$ field fixed at various values during the fitting routine and plotted as the difference in $\chi^2_{red}$ from the best fit value, is shown in row 2. The bootstrap distribution for each fit is shown in row 3. Rows 4 and 5 are the $\chi^2_{red}$ surfaces and bootstrap distributions for the inhomogeneity in each $B_1$ field. The $B_1$ fields and errors obtained from the bootstrap distributions are indicated above each panel in row 1, while the corresponding $I_{B1}$ magnitudes and errors are indicated above panels in row 4. $I_{B1}$ values were obtained only for $B_1$ fields larger than 5 Hz.

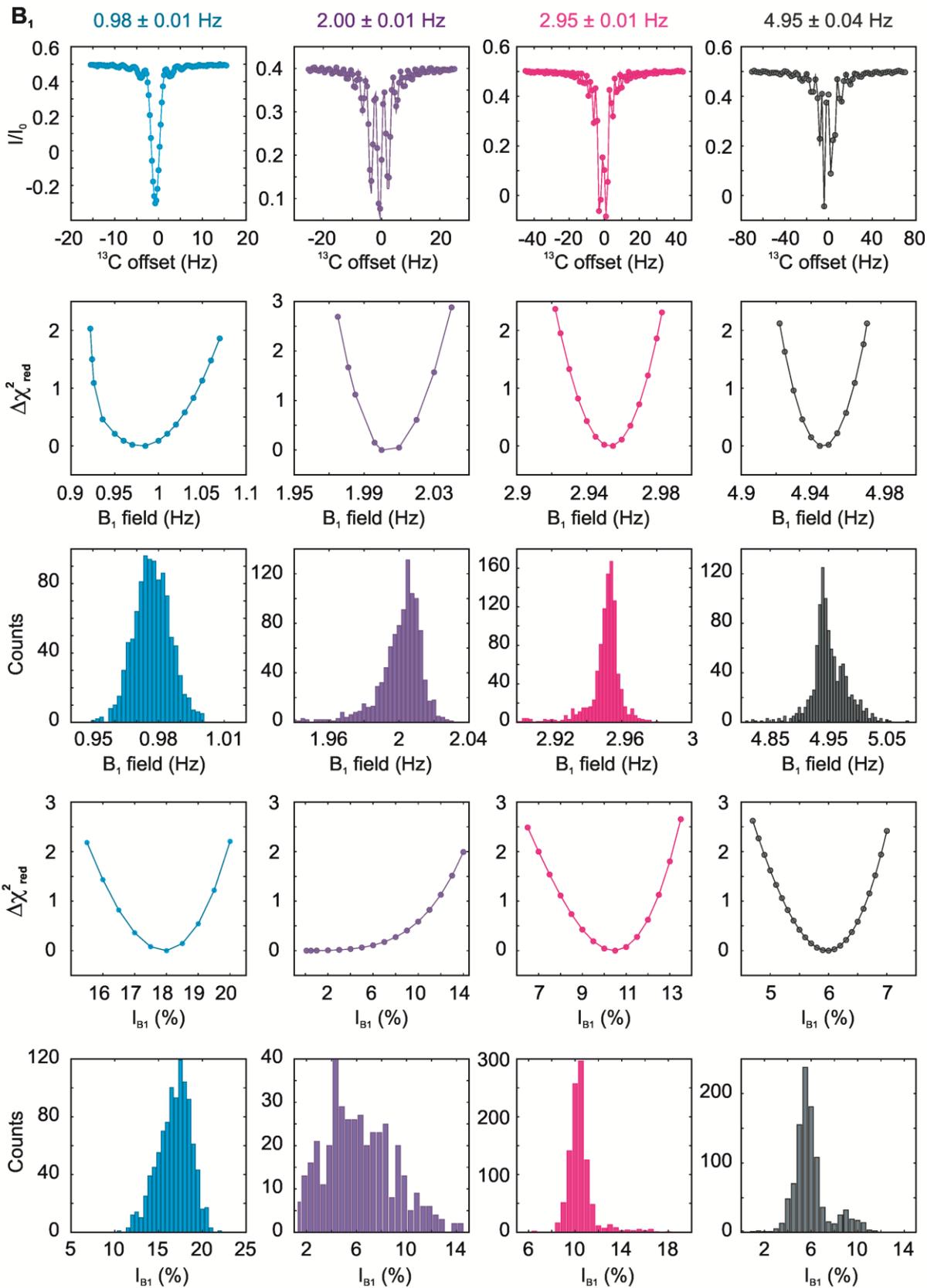

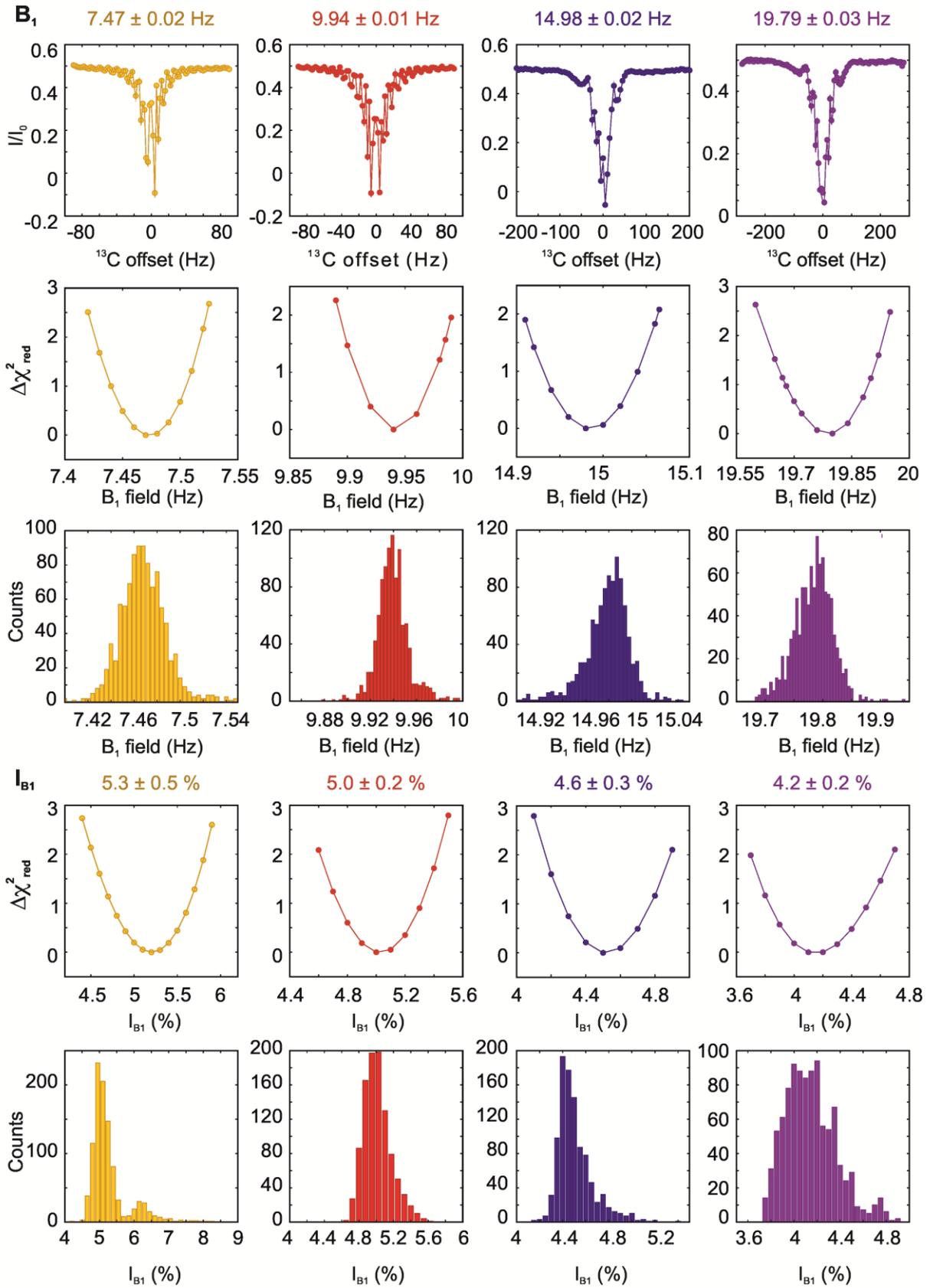

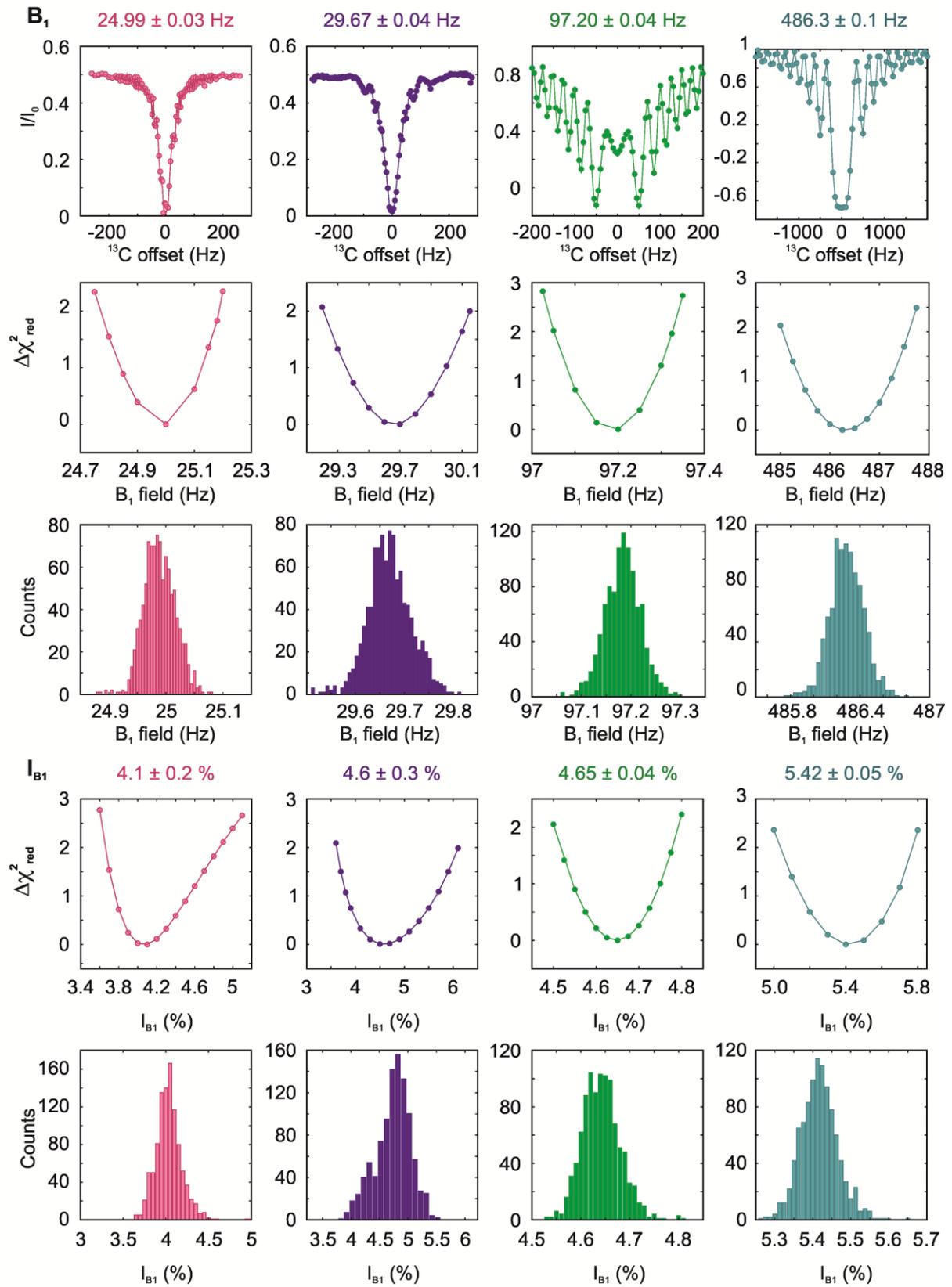

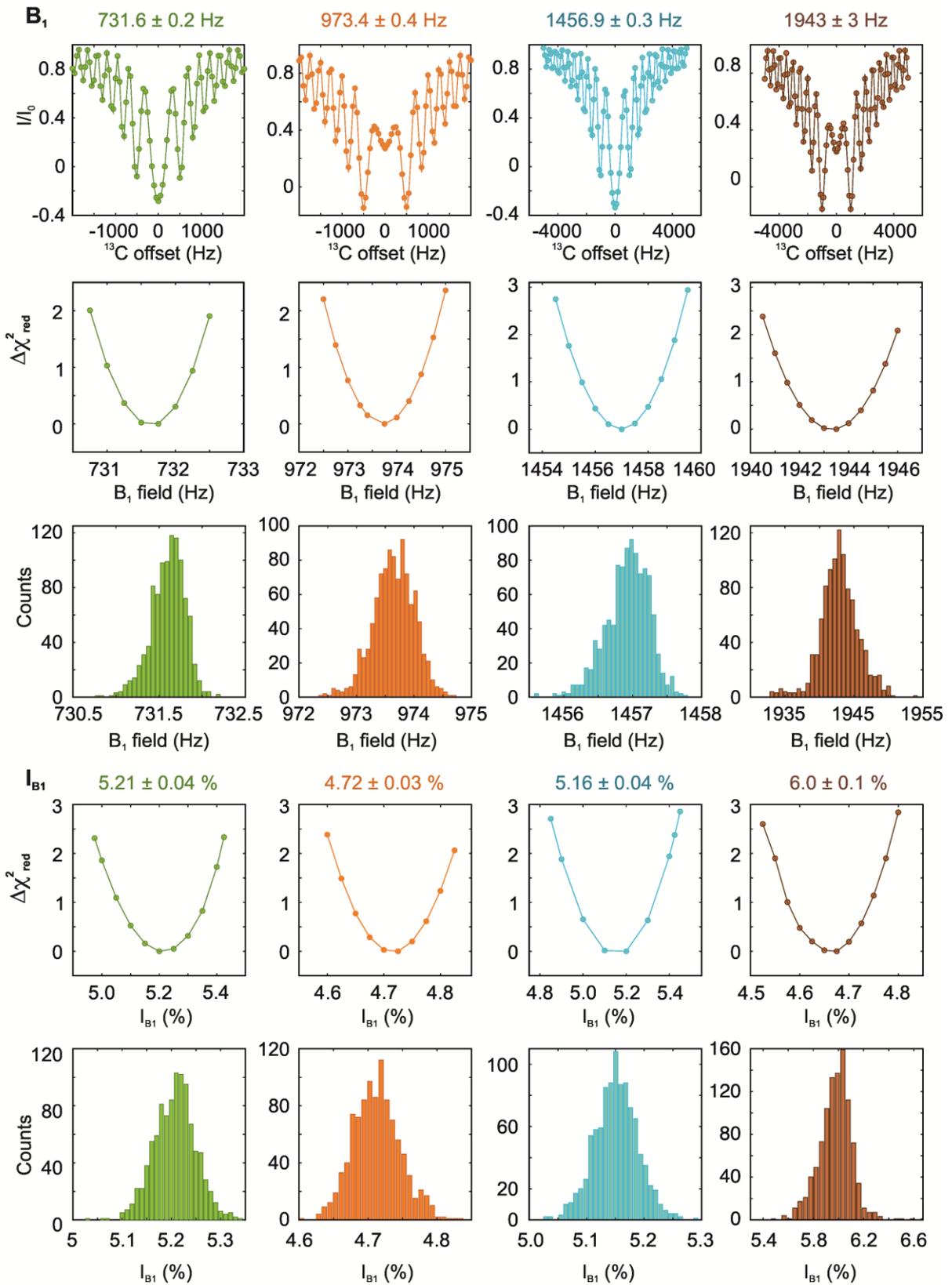

**Figure S2.** Overlay of simulated CONDENZ profiles for a single $^{13}$C spin (open blue circles) and an AMX spin system (A = $^{13}$C, M = X = $^{1}$H, filled red squares) for B$_1$ field strengths ranging from 0.5 - 10 Hz applied on-resonance to the $^{13}$C spin in both cases. For the AMX spin system, a 4 kHz DIPSI-2 composite pulse decoupling train is applied on-resonance to the M spin during $t_{nut}$, while the X spin is off-resonance by 1.3 kHz. Scalar coupling constants within the AMX spin system were set to $^{1}J_{AM}$ = 169 Hz, $^{2}J_{AX}$ = 7 Hz, and $^{3}J_{MX}$ = 4 Hz. The parameters for the AMX spin system were chosen based on the assignment of the sucrose anomeric $^{13}$C as the A spin. The parameters common to both simulations are: $t_{nut}$ = 0.4 s, R$_1$ = 1.7 s$^{-1}$, R$_2$ = 2.0 s$^{-1}$ and I$_{B1}$ = 10 %. It can be seen clearly from the CONDENZ profiles that the single spin-1/2 (open blue circles) and AMX data (filled red squares) overlay perfectly, demonstrating that the AMX spin system in the presence of DIPSI-2 composite pulse decoupling can be very well approximated to a single-spin-1/2 system for the purposes of quantifying CONDENZ data. The B$_1$ field strengths recovered by fitting the AMX CONDENZ profiles to the single-spin-1/2 Bloch equations (main manuscript, Eq. 9) are tabulated in Table S2 and agree excellently with the input values, consistent with the perfect overlay between the single spin and AMX three-spin data shown in this figure.

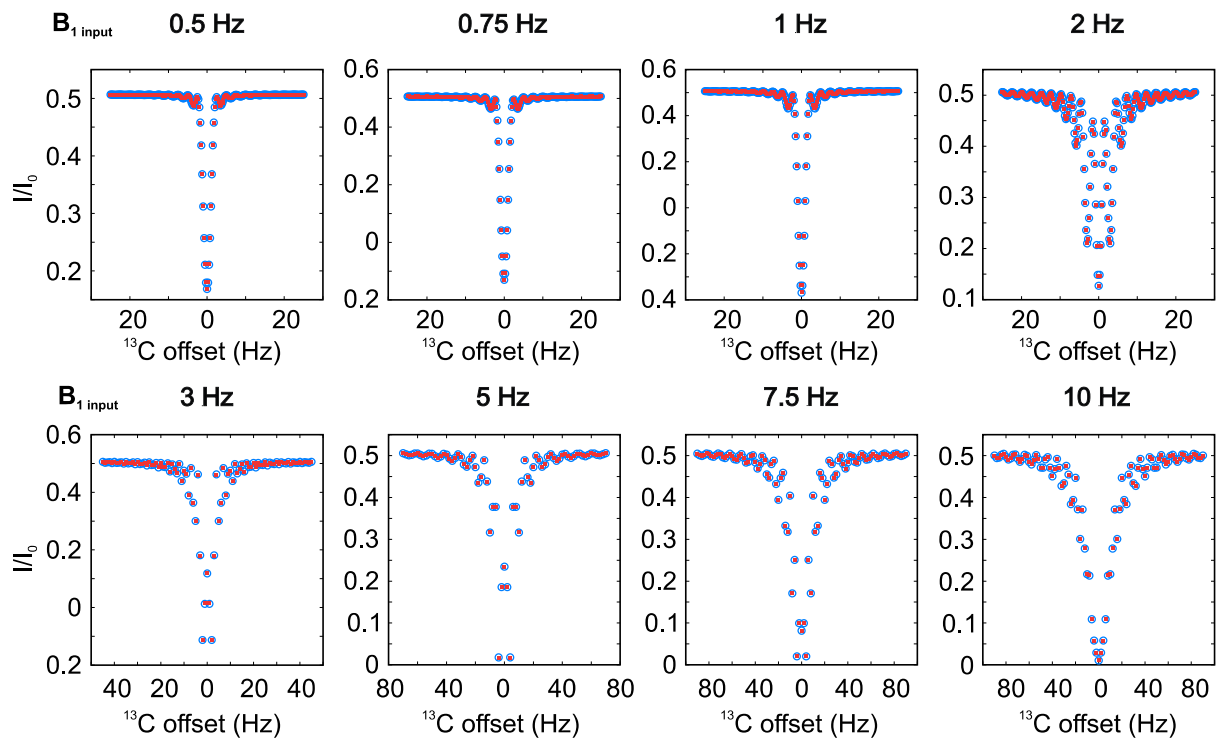

**Figure S3**. CONDENZ profiles (row 1) acquired on $^{15}$N$^\varepsilon$-labeled Trp at various B$_1$ fields ranging from 1 - 50 Hz. Solid lines are fits of the data to the single-spin-1/2 Bloch equations (main manuscript, Eq. 9). The $\chi^2_{red}$ surface for each fit, plotted as the difference in $\chi^2_{red}$ from the best fit value, evaluated by keeping the B$_1$ field fixed at various values during the fitting routine, is shown in row 2, while the bootstrap distribution for each fit is shown in row 3. Rows 4 and 5 are the $\chi^2_{red}$ surfaces and bootstrap distributions for the inhomogeneity in each B$_1$ field. The B$_1$ fields and errors obtained from the bootstrap distributions are indicated above each panel in row 1, while the corresponding I$_{B1}$ magnitudes and errors are indicated above panels in row 4. I$_{B1}$ values were determined only for B$_1$ fields larger than 5 Hz.

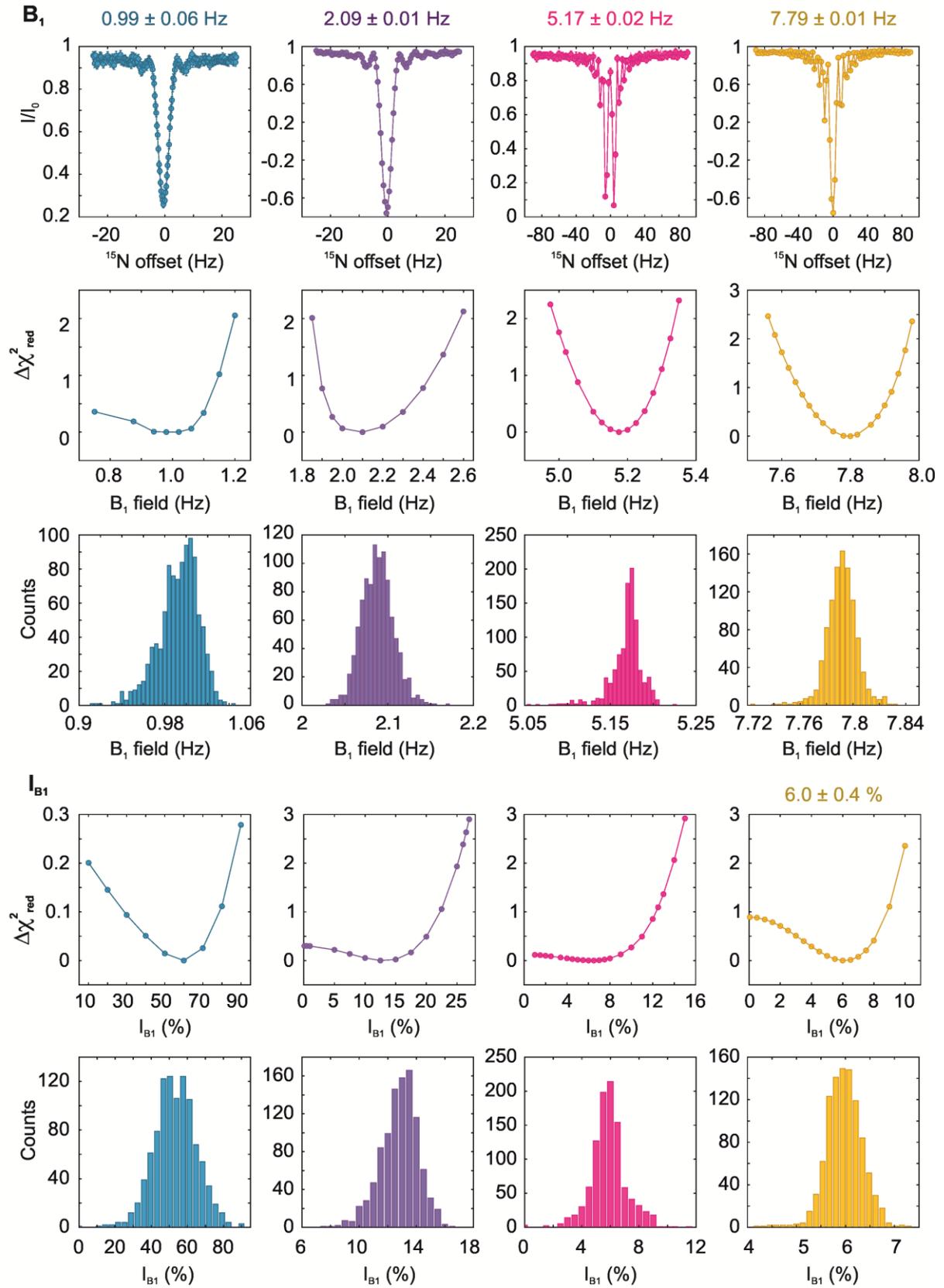

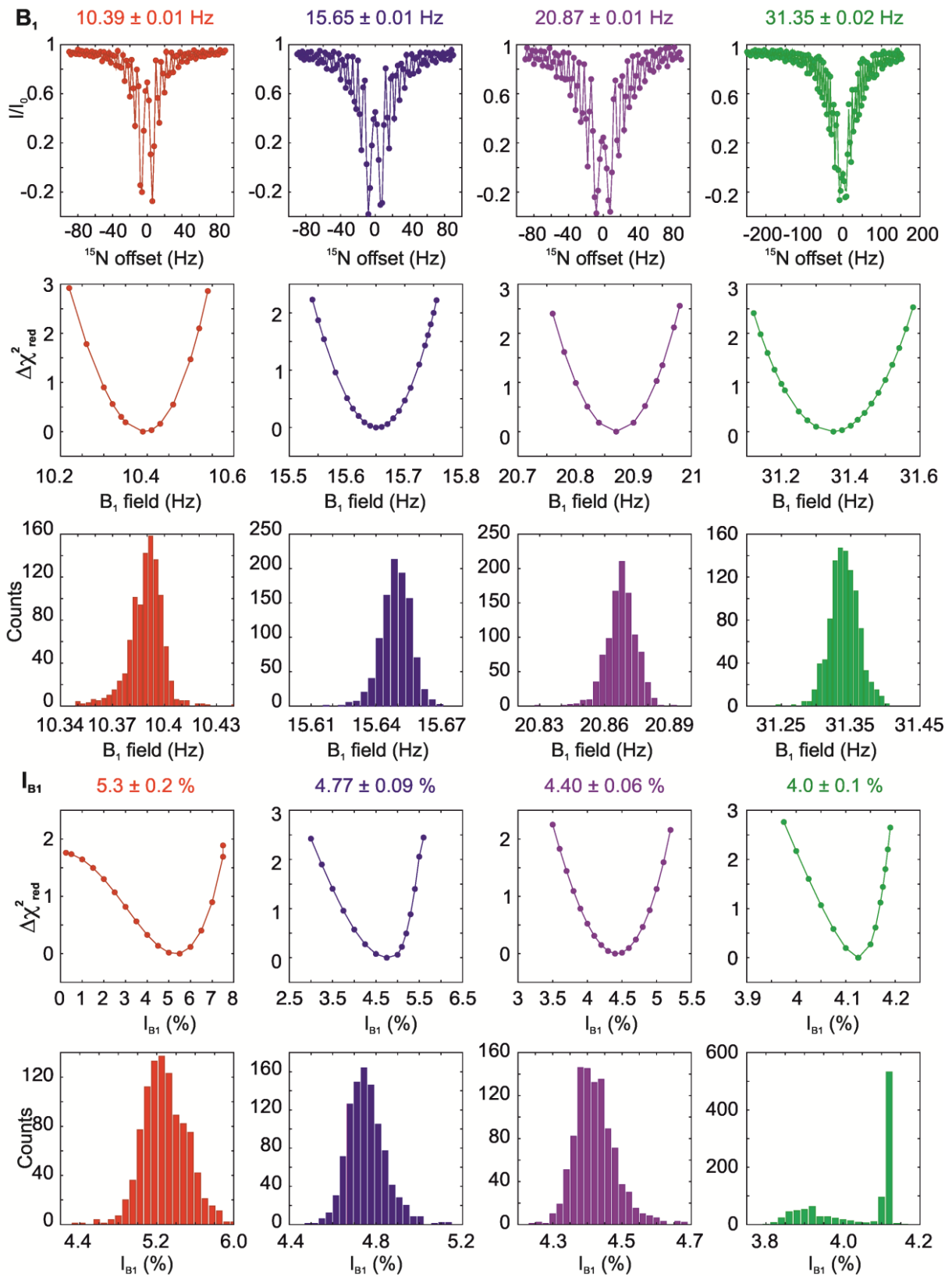

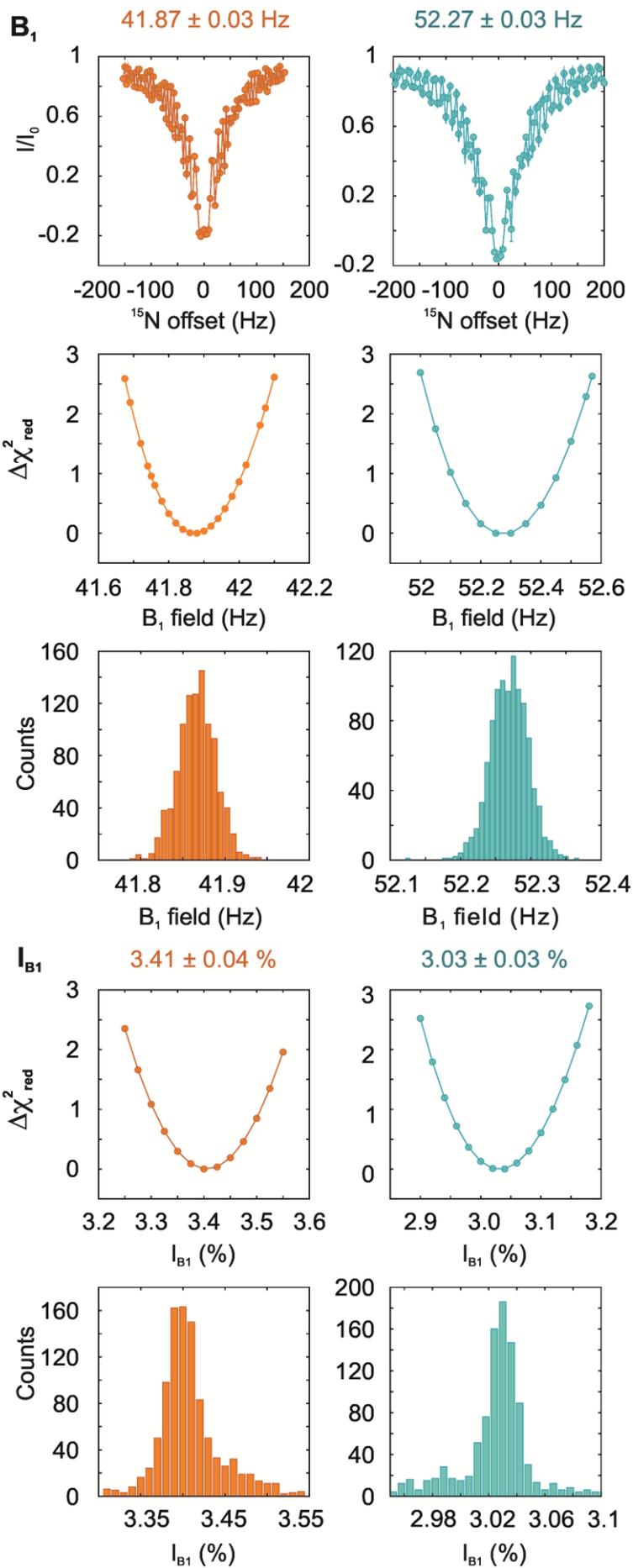

# Supporting Tables

**Table S1**. The offset sweep width, offset spacing and nutation period ($t_{nut}$) which were used to acquire the CONDENZ profiles in Figure 1 and Figure S1.

| $B_{1\,setting}$ (Hz) | $t_{nut}$ (ms) | Sweep width (Hz) | Offset spacing (Hz) |
|---|---|---|---|
| 1 | 400 | -15.5 to 15.5 | 0.25 |
| 2 | | -25 to 25 | 0.5 |
| 3 | | -45 to 45 | 1 |
| 5 | | -70 to 70 | 2 |
| 7.5 | | -90 to 90 | 2 |
| 10 | | -90 to 90 | 2 |
| 15 | | -200 to 200 | 5 |
| 20 | | -280 to 280 | 5 |
| 25 | | -260 to 260 | 6 |
| 30 | | -280 to 280 | 5 |
| 100 | 50 | -200 to 200 | 5 |
| 500 | 5 | -2000 to 2000 | 50 |
| 750 | | -2000 to 2000 | 50 |
| 1000 | | -2000 to 2000 | 50 |
| 1500 | 2.5 | -5000 to 5000 | 100 |
| 2000 | | -5000 to 5000 | 100 |

**Table S2**: RF field amplitudes ($B_1$), $B_1$ inhomogeneities ($I_{B1}$), $R_1$ and $R_2$ estimated by fitting CONDENZ data simulated for an AMX spin system (A = $^{13}$C, M = X = $^1$H) with scalar coupling values $^1J_{AM}$ = 169 Hz, $^2J_{AX}$ = 7 Hz, and $^3J_{MX}$ = 4 Hz. The A and M spins are on-resonance to the CW and 4 kHz DIPSI-2 RF fields respectively, while the X spin is 1.3 kHz off-resonance. Input values used were: $R_{1\ input}$ = 1.7 s$^{-1}$, $R_{2\ input}$ = 2.0 s$^{-1}$, $I_{B1\ input}$ = 10 %. The CW $B_1$ field was varied in the simulations over a range of values from 0.5 - 10 Hz. The simulated data were then modeled using the Bloch equations for a single-spin-1/2 system (main manuscript, Eq. 9) to derive the fit parameters. The simulated data are shown in Figure S2.

| $B_{1\ input}$ (Hz) | $B_{1\ fit}$ (Hz) | $I_{B1\ fit}$ (%) | $R_{1\ fit}$ (s$^{-1}$) | $R_{2\ fit}$ (s$^{-1}$) |
|---|---|---|---|---|
| 0.5 | 0.500 | 9.45 | 1.70 | 2.01 |
| 0.75 | 0.750 | 9.45 | 1.70 | 2.01 |
| 1.0 | 0.999 | 9.94 | 1.70 | 2.01 |
| 2.0 | 1.999 | 9.99 | 1.70 | 2.01 |
| 3.0 | 2.998 | 10.01 | 1.70 | 2.01 |
| 5.0 | 4.997 | 10.00 | 1.70 | 2.01 |
| 7.5 | 7.495 | 10.00 | 1.70 | 2.01 |
| 10 | 9.994 | 10.00 | 1.70 | 2.00 |

**Table S3**. $^{15}$N RF fields and inhomogeneities measured using $^{15}$N$^{\varepsilon}$-tryptophan. All CONDENZ profiles were measured using a $t_{nut}$ of 200 ms, and the offset frequency and spacing were similar to the values used for sucrose (Table S1).

| $B_{1\,setting}$ (Hz) | $B_{1\,CONDENZ}$ (Trp$_{int}$, Hz) | $I_{B1}$ (%) |
|---|---|---|
| 1 | 0.99 ± 0.06 | ND |
| 2 | 2.09 ± 0.01 | ND |
| 5 | 5.17 ± 0.02 | ND |
| 7.5 | 7.79 ± 0.01 | 6.0 ± 0.4 |
| 10 | 10.39 ± 0.01 | 5.3 ± 0.2 |
| 15 | 15.65 ± 0.01 | 4.77 ± 0.09 |
| 20 | 20.87 ± 0.01 | 4.40 ± 0.06 |
| 30 | 31.35 ± 0.02 | 4.0 ± 0.1 |
| 40 | 41.87 ± 0.03 | 3.41 ± 0.04 |
| 50 | 52.27 ± 0.03 | 3.03 ± 0.03 |

ND: Not determined

**Table S4**: A comparison of RF fields measured on $^{15}$N nuclei using a) the on-resonance nutation experiment on the Gln62 amide resonance of U-$^{15}$N ubiquitin, b) the CONDENZ approach with $^{15}$N$^{\varepsilon}$-tryptophan as an internal standard in the sample containing U-$^{15}$N ubiquitin, c) CONDENZ profiles with $^{15}$N$^{\varepsilon}$-tryptophan as an external standard in a separate sample but without changing the tuning and matching capacitor settings after measuring the $B_1$ field on the U-$^{15}$N ubiquitin sample, and d) CONDENZ data on a separate $^{15}$N$^{\varepsilon}$-tryptophan sample after tuning and matching. $B_{1\ nutation}$ values are reported to the same number of decimal places as the $B_{1\ CONDENZ}$ values to facilitate comparison.

| $B_{1\ nutation}$ (ubiquitin, Hz) | $B_{1\ CONDENZ}$ (Trp$_{int}$, Hz) | $B_{1\ CONDENZ}$ (Trp$_{ext}$, Hz) | $B_{1\ CONDENZ}$ (Trp$_{ext}$, after tuning/matching, Hz) |
|---|---|---|---|
| 0.86 | 0.99 ± 0.06 | 1.03 ± 0.03 | 1.08 ± 0.01 |
| 2.08 | 2.09 ± 0.01 | 2.07 ± 0.02 | 2.22 ± 0.02 |
| 5.19 | 5.17 ± 0.02 | 5.19 ± 0.01 | 5.24 ± 0.02 |
| 10.47 | 10.39 ± 0.01 | 10.44 ± 0.01 | 10.51 ± 0.01 |
| 15.67 | 15.65 ± 0.01 | 15.70 ± 0.01 | 15.80 ± 0.01 |
| 20.93 | 20.87 ± 0.01 | 20.99 ± 0.02 | 21.12 ± 0.01 |
| 31.36 | 31.35 ± 0.02 | 31.42 ± 0.02 | 31.69 ± 0.02 |
| 41.82 | 41.87 ± 0.03 | 41.99 ± 0.01 | 42.26 ± 0.02 |

**Table S5**: Summary of RF fields measured using the CONDENZ method on the methyl carbon of $^{13}CH_3$ α-ketobutyric acid ($\varpi$ = 6.5 ppm), the anomeric $^{13}C$ of sucrose ($\varpi$ = 92.2 ppm), the aromatic $^{13}C$ of benzaldehyde ($\varpi$ = 135.5 ppm, the carbon at the ortho position to the aldehyde group) and the aldehydic $^{13}C$ of benzaldehyde ($\varpi$ = 196.5 ppm).

| $B_{1\ setting}$ (Hz) | $B_1$ (Hz) $^{13}C^\delta$ = 6.5 ppm | $B_1$ (Hz) $^{13}C^\delta$ = 92.2 ppm | $B_1$ (Hz) $^{13}C^\delta$ = 135.5 ppm | $B_1$ (Hz) $^{13}C^\delta$ = 196.5 ppm |
|---|---|---|---|---|
| 2  | 1.72 ± 0.02  | 2.1 ± 0.2    | 2.03 ± 0.01  | 1.98 ± 0.01  |
| 5  | 5.11 ± 0.01  | 5.13 ± 0.02  | 5.11 ± 0.01  | 5.09 ± 0.01  |
| 10 | 10.18 ± 0.02 | 10.32 ± 0.03 | 10.21 ± 0.02 | 10.17 ± 0.01 |
| 20 | 20.31 ± 0.01 | 20.63 ± 0.02 | 20.43 ± 0.01 | 20.21 ± 0.03 |